\documentclass[a4paper,20pt]{article}
    \usepackage[T1]{fontenc}
    \usepackage{graphicx}
    \usepackage{epsfig}
    \usepackage{amsmath}
    \usepackage{amsfonts}
    \renewcommand{\abstract}{}
\begin{document}
\makeatletter
\renewcommand{\@oddhead}{\textit{} 
\hfil 
\textit{R. Szcz{\c{e}}{\`s}niak, M.W. Jarosik, M. Szcz{\c{e}}{\`s}niak, A. Durajski}}
\renewcommand{\@evenfoot}{\hfil \thepage \hfil}
\renewcommand{\@oddfoot}{\hfil \thepage \hfil}
\fontsize{11}{11} \selectfont

\title{The Final State of the Thermal Evolution of Free-Floating Giant Planet}
\author{\textsl{R. Szcz{\c{e}}{\`s}niak$^{1}$, M.W. Jarosik$^{1}$, 
                M. Szcz{\c{e}}{\`s}niak$^{2}$, A. Durajski$^{1}$}}
\date{}
\maketitle
\begin{center} {\small 
$^{1}$Institute of Physics, Cz{\c{e}}stochowa University of Technology, Al. Armii Krajowej 19, 42-200 Cz{\c{e}}stochowa, Poland\\
$^{2}$Ul. S{\l}oneczna 44, 34-700 Rabka-Zdr{\'o}j, Poland\\
szczesni@mim.pcz.czest.pl}
\end{center}

\begin{abstract}
In the paper, we have discussed the problem of the existence of the metallic hydrogen in the superconducting state inside the cold giant planet. We have shown that the {\it cold} planet represents the final state of the thermal evolution of free-floating giant planet.
\end{abstract}
\section*{Introduction}
\indent \indent 
In the year of $1935$ Wigner and Huntington predicted that under the influence of high pressure ($p$) the molecular phase of solid hydrogen passes into a metallic phase \cite{Wigner}. Recently, the theoretical calculations fixed the metallization pressure for the solid hydrogen at $400$ GPa \cite{Stadele}. In the case of the liquid hydrogen, Nellis {\it et al.} inform about the observation of the insulator-metal transition for pressure of $140$ GPa and the temperature $3000$ K \cite{Weir}. This result confirms the fact  that the liquid hydrogen in the metallic state exists inside Jupiter and Saturn \cite{Guillot}. We notice that the hydrogen phase diagram from the planetary science aspect is in detail discussed by Fortney and Hubbard in \cite{Fortney1}. In the year of 1968 Ashcroft suggested the possibility of occurrence of  additional phase transition in the solid hydrogen; from the metallic to the superconducting state \cite{Ashcroft}. The obtained results concerning superconductivity in the solid metallic hydrogen can shed light on the internal structure of the hypothetical cold giant planets. 

We define the {\it cold} planets as a class of  planets whose mass is close to the mass of Saturn or exceeds it. Additionally, they have the characteristics: (i) The {\it cold} planets are primarily composed of the hydrogen. (ii)  Some part of the solid metallic hydrogen is in the superconducting state. 

The liquid metallic hydrogen inside the planet like Saturn exists between 200 GPa and 1000 GPa \cite{Stevenson}. So, only the low-pressure superconducting state one can consider 
in  the solid hydrogen for low values of  the temperature. In the model calculations the best choice represents the planet that possessess the Jupiter's mass, where the liquid metallic hydrogen is induced between 200 GPa and 4000 GPa \cite{Stevenson}. 

In the paper we will calculate the maximum surface temperature of the {\it cold} Jupiter. We will consider the high- and low-pressure superconducting state inside the planet 
($p=2000$ GPa or $450$ GPa). We notice that these states are characterized by the maximum values of  the critical temperature ($T_{C}$). Additionally, we will show that {\it cold} planet represents the final state of the thermal evolution of free-floating giant planet.  

\section*{The critical temperature for the metallic hydrogen: $p=2000$ GPa}
\indent\indent 
The superconducting properties of the metallic hydrogen for $p=2000$ GPa are determined by the Eliashberg equations \cite{eliashberg}:
\begin{equation}
\label{r1}
\Delta_lZ_{l}=\frac{\pi}{\beta} \sum_{m=1}^{M}
\frac{\left[K^{+}\left(l,m\right)-2\mu\left(m\right)\right]} 
{\sqrt{\omega_m^2+\Delta_m^2}}
\Delta_{m}, 
\end{equation}
\begin{equation}
\label{r2}
Z_l=1+\frac {\pi}{\beta }\sum_{m=1}^{M}
\frac{K^{-}\left(l,m\right)}
{\sqrt{\omega_m^2+\Delta_m^2}}\frac{\omega_m}{\omega _l},
\end{equation}
where $\Delta_{l}\equiv\Delta\left(i\omega_{l}\right)$ and $Z_{l}\equiv Z\left(i\omega_{l}\right)$ are the superconducting order parameter and the wave function renormalization factor respectively. The symbol $\omega_{l}\equiv \left(\pi / \beta\right)\left(2l-1\right)$ denotes the Matsubara frequencies; $\beta\equiv\left(k_{B}T\right)^{-1}$, where $k_{B}$ is the Boltzmann constant. 
The functions $K^{\pm}\left(l,m\right)$ are defined by the formula: 
$K^{\pm}\left(l,m\right)\equiv K\left(l-m\right)\pm K\left(l+m-1\right)$,
where $K\left(l-m\right)$ is the pairing kernel:
\begin{equation}
\label{r3}
K\left(l-m\right)\equiv\int_0^{\Omega_{\rm{max}}}d\Omega\frac{\alpha^{2}F\left(\Omega\right) 
2\Omega}{\left(\omega_l-\omega_m\right)
^2+\Omega ^2}.
\end{equation}

The Eliashberg function ($\alpha^{2}F\left(\Omega\right)$) for $p=2000$ GPa was determined in \cite{Maksimov2}; the maximum phonon frequency ($\Omega_{\rm{max}}$) is equal to $570$ meV.

The influence of in-between electron Coulomb interaction on the superconducting state is modeled by the function: 
$\mu\left(m\right)\equiv \mu^{*}\Theta\left(\omega_{\rm{c}}-\left|\omega_{m}\right|\right)$,
 where $\mu^{*}$ is the Coulomb pseudopotential. The symbol $\Theta$ denotes the Heaviside unit function and $\omega_{c}$ is the phonon cut-off frequency ($\omega_{c}=3\Omega_{\rm{max}}$).

By using the Eliashberg equations we have shown in the paper \cite{Szczesniak1} that the dependency of the critical temperature on the Coulomb pseudopotential can be precisely parameterized by the {\it modified} Allen-Dynes formula \cite{allen}:
\begin{equation}
\label{r4}
k_{B}T_{C}=f_{1}f_{2}\frac{\omega_{{\rm ln}}}{1.2}\exp\left[\frac{-1.04\left(1+\lambda\right)}
{\lambda-\mu^{*}\left(1+0.62\lambda\right)}\right],
\end{equation}
where the electron-phonon coupling constant ($\lambda$) and the logarithmic phonon frequency ($\omega_{{\rm ln}}$) are defined by: 
\begin{equation}
\label{r5}
\lambda\equiv 2\int^{\Omega_{\rm{max}}}_{0}d\Omega\frac{\alpha^{2}F\left(\Omega\right)}{\Omega},
\end{equation}
\begin{equation}
\label{r6}
\omega_{{\rm ln}}\equiv \exp\left[\frac{2}{\lambda}
\int^{\Omega_{\rm{max}}}_{0}d\Omega\frac{\alpha^{2}F\left(\Omega\right)}
{\Omega}\ln\left(\Omega\right)\right].
\end{equation}
In the considered case $\lambda=7.32$ and $\omega_{{\rm ln}}=89.2$ meV.
The functions $f_{1}$, $f_{2}$ have the form \cite{allen}:
\begin{equation}
\label{r7}
f_{1}\equiv\left[1+\left(\frac{\lambda}{\Lambda_{1}}\right)^{\frac{3}{2}}\right]^{\frac{1}{3}},
\qquad
f_{2}\equiv 1+\frac
{\left(\frac{\sqrt{\omega_{2}}}{\omega_{\rm{ln}}}-1\right)\lambda^{2}}
{\lambda^{2}+\Lambda^{2}_{2}},
\end{equation}
where: 
\begin{equation}
\label{r8}
\omega_{2}\equiv 
\frac{2}{\lambda}
\int^{\Omega_{\rm{max}}}_{0}d\Omega\alpha^{2}F\left(\Omega\right)\Omega
\end{equation}
and $\omega_{2}=25.942$ eV. 
Finally:
\begin{equation}
\label{r9}
\Lambda_{1}\equiv 2.41\left(1-1.285\mu^{*}\right), \qquad 
\Lambda_{2}\equiv 3.15\left(1+5.513\mu^{*}\right)\left(\frac{\sqrt{\omega_{2}}}{\omega_{\rm{ln}}}\right).
\end{equation}

The value of $T_{C}$ can be easily calculated on the basis of  the formula \ref{r4}. In particular, for $\mu^{*}\in\left(0.1, 0.5\right)$ the critical temperature changes the value from $631$ K to $413$ K.

\section*{The characteristic of the {\it cold} Jupiter}
\indent \indent       
Below, we calculate the maximum surface temperature of the {\it cold} Jupiter $\left[T_{s}\right]_{\rm{max}}$ (the temperature at the pressure $p_{\rm{s}}$=100 kPa).
     In the first step, we determine the relationship between the surface temperature 
and the internal temperature $T_{\rm{in}}$ (the temperature at the pressure $p_{\rm{in}}> p_{\rm{s}}$). By using the adiabatic model we obtain \cite{Stevenson}: 
\begin{equation}
\label{r11} 
T_{\rm{s}}=\left(\frac{p_{\rm{s}}}{p_{\rm{in}}}\right)^{\Gamma} T_{\rm{in}},
\end{equation}
where the  adiabatic index $\Gamma$ equals the Jupiter's index;  $\Gamma =0.275$. 
The parameter $\Gamma$ has been calculated under the assumption that the surface temperature of  the Jupiter equals $170$ K and $T_{\rm{in}}=21000$ K for $p_{\rm{in}}=4000$ GPa \cite{Stevenson}.

     Next, for the high-pressure superconducting state we take: $p_{\rm{in}}=2000$ GPa and $T_{C}\in\left(413, 631\right)$. To describe the low-pressure case we use: $p_{\rm{in}}=450$ GPa and $T_{C}=242$ K \cite{Cudazzo}. We notice that this state is characterized by the maximum value of $T_{C}$ and its properties are obtained in the framework of the recently introduced density functional theory of superconductivity.

\begin{figure}[t]%
\centering
\includegraphics*[scale=0.15]{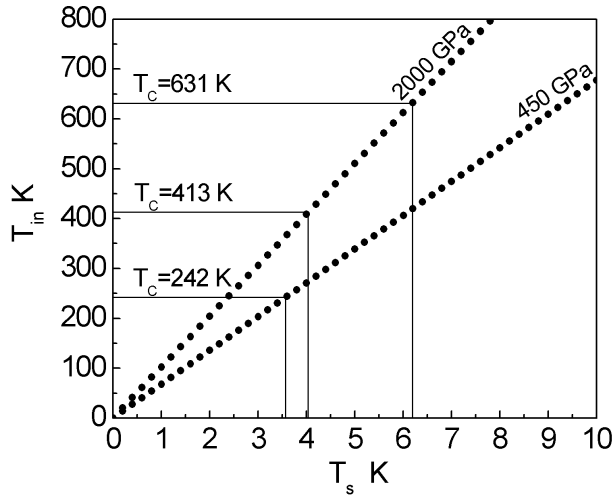}
\caption{ The dependence of  the internal temperature on the surface temperature for considered values of $p_{\rm{in}}$. The horizontal lines denote the values of the critical temperature.}
\label{f1}
\end{figure}

     The dependence of  the internal temperature on the surface temperature is shown 
in the figure \ref{f1}. The horizontal lines correspond to the values of the critical temperature. 
On the basis of  the presented results, we conclude that the maximum surface temperature of the {\it cold} Jupiter is situated between 4 K and 6.2 K.           

%
%

%
\begin{figure}[t]%
\centering
\includegraphics*[scale=0.15]{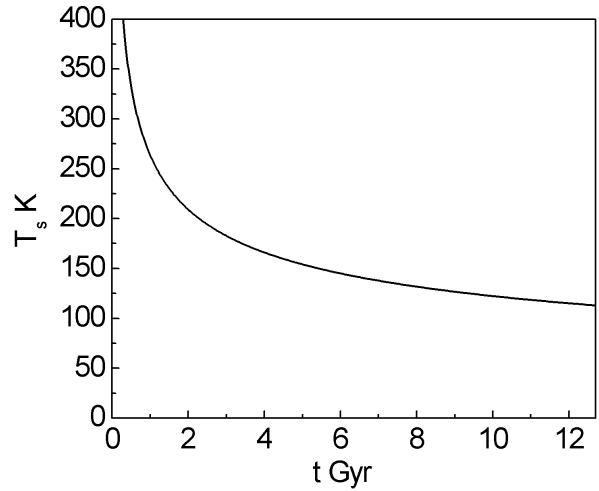}
\caption{The dependence of the surface temperature on the time.}
\label{f2}
\end{figure}
%

     The superconducting state inside the Jupiter can not exist because the value of the Jupiter's surface temperature is too high. This fact is connected with Jupiter's young age ($4.7$ Gyr) \cite{Fortney2}. 
     In the last step, we calculate the surface temperature of the old free-floating giant planet. 
We assume that the planet's age is equal to $12.7$ Gyr (the age of the oldest known planet) \cite{Thorsett} and the maximum temperature rises during the formation to $10^{5}$ K \cite{Stevenson}.

     The evolutionary model of the giant planet was first presented by Hubbard \cite{Hubbard}.  The thermal history calculations for non-rotating planet of the Jupiter's radius $R_{\rm{J}}$ and the Jupiter's mass $M_{\rm{J}}$ can be done by using the equation:
\begin{equation}
\label{r12} 
4\pi R_{\rm{J}}^{2}\sigma T^{4}_{\rm{eff}}=L\left(M_{\rm{J}}, t, X\right),
\end{equation}
where $\sigma$ is the Stefan-Boltzmann constant, $T_{\rm{eff}}$ is the planet's effective temperature, $L$ is the planet's luminosity and $t$ is the planet's age. The symbol $X$ denotes the specified composition of the various mass fractions. In particular, we assume that 
the planet has the core of the mass $10M_{\rm{Earth}}$ and its thermal evolution is independent of the separation of the helium from the metallic hydrogen. 
     
     The figure \ref{f2} shows our result. We see that in the Universe the {\it cold} Jupiter can not yet exist because the surface temperature for $t=12.7$ Gyr is too high ($113$ K). We obtain that the superconducting state will be induced inside the planet for  $t\simeq 10^{5}$ Gyr. So, the {\it cold} planet represents the final state of the thermal evolution of the free-floating giant planet.

\section*{Conclusions}
\indent \indent 
We have calculated the maximum surface temperature of the {\it cold} Jupiter whose internal structure is close to structure of Jupiter. We have shown that  $\left[T_{s}\right]_{\rm{max}}$ is situated between $4$ K and $6.2$ K. We have proved that the {\it cold} planet represents the final state of the thermal evolution of the free-floating giant planet.

\section*{Acknowledgement}
\indent \indent Authors would like to thank Dr Bogdan Wszo{\l}ek for given organizational aid. 

\end{document}